# Assessing the Impact of Model Assumptions in Network Meta-Regression: A Simulation Study

**Nana-adjoa Kwarteng[1], Guido Schwarzer[1], Adriani Nikolakopoulou[1,2], Theodoros Evrenoglou[1]**

[1]Institute of Medical Biometry and Statistics, Faculty of Medicine and Medical Center, University of Freiburg, Freiburg im Breisgau, Germany

[2]Department of Hygiene, Social-Preventive Medicine and Medical Statistics, School of Medicine, Aristotle University of Thessaloniki, Thessaloniki, Greece

**Corresponding Author:**

Nana-adjoa Kwarteng

Institute of Medical Biometry and Statistics

Faculty of Medicine and Medical Center - University of Freiburg

Stefan-Meier-Str. 26, 79104 Freiburg im Breisgau, Germany

Email: Nana.Kwarteng@uniklinik-freiburg.de

## Abstract

Network meta-regression (NMR) extends network meta-analysis (NMA) by synthesizing evidence on multiple treatments while adjusting for potential effect modifiers. By accounting for effect modification, NMR can reduce between-study heterogeneity and improve the validity of relative treatment effects, providing insight regarding characteristics impacting treatment performance. However, choosing between available NMR models is complex, as each model addresses a similar, but unique research question, and the performance of available NMR models under varying network structures, between-study heterogeneity, and interaction assumptions remains unclear. We evaluated the consequences of model misspecification in a simulation study of 120 evidence-network scenarios designed to reflect potential complications in evidence networks introduced by trial design, heterogeneity levels, and interaction assumptions. We compared the standard interaction-free NMA model with four NMR parameterizations differing in across-comparison interaction assumptions (common vs. independent interactions) and interaction consistency assumptions (with or without consistency). Standard NMA models generally overestimated treatment effects when effect modification was present. NMR models with independent across-comparison interactions maintained appropriate confidence interval coverage in dense networks generated with their corresponding consistency assumptions. However, their coverage deteriorated in sparse networks with between-study heterogeneity. Models assuming consistent interactions are advantageous in networks with multi-arm studies. Ignoring effect modification in NMA can lead to biased treatment effect estimates. When effect modification is anticipated, thoughtful alignment between network structure and NMR assumptions can reduce bias and misleading precision, supporting more reliable conclusions in medical decision making. These findings provide guidance for planning and interpreting network meta-analyses and meta-regressions in complex networks.

# 1. Introduction

Network meta-analysis (NMA) simultaneously compares multiple treatments using evidence from several individual studies (Ades et al., 2024; Borenstein, 2009; Caldwell et al., 2010; Deeks et al., 2019; Dias et al., 2018; Higgins & Whitehead, 1996; Lu & Ades, 2004; Lumley, 2002; Nixon et al., 2007; Salanti, 2012; White et al., 2012). Synthesis of multiple studies involves the estimation of between-study heterogeneity, which can arise from variation in treatment effects across studies as well as differences in effect modifiers (Dias et al., 2018; Dias, Sutton, et al., 2013; Donegan et al., 2019; Higgins et al., 2019; Thompson & Sharp, 1999). If the distribution of an effect-modifier varies between studies or comparisons, standard NMA treatment effect estimates may be biased and lead to increased estimates of between-study heterogeneity (Cooper et al., 2009; Phillippo et al., 2020). Network meta-regression (NMR) addresses the potential between-study heterogeneity introduced by effect modifiers by modeling treatment-by-covariate interactions, enabling the assessment of effect modification across a treatment network (Cooper et al., 2009). In cases where the interaction substantially accounts for residual heterogeneity across studies, interpreting NMR parameters may provide clinically relevant insights (Donegan et al., 2015, 2017). Ideally, researchers examining patient-level effect modifiers such as age and gender should use individual participant data (IPD) for NMR (Riley et al., 2008). However, IPD are often unavailable due to obstacles in data acquisition, and, in such cases, NMR using aggregate data (AD) provides a practical alternative for investigating between-study heterogeneity (Chaimani, 2015; Polanin & Williams, 2016; Schmid et al., 2004; Van Der Worp et al., 2023).

Implementation of NMR requires choices of modeling assumptions that can influence both the estimation and interpretation of treatment and interaction effects (Dias et al., 2018; Dias, Sutton, et al., 2013; Donegan et al., 2019). These assumptions include whether consistency is imposed across interaction terms and how interactions are structured across treatment comparisons, which can be independent, common, or exchangeable (Dias et al., 2018; Donegan et al., 2019). The consistency assumption for interaction terms parallels the standard NMA consistency assumption for treatment effects. All effects are first expressed as basic pairwise comparisons relative to a single network reference treatment, and remaining comparisons are obtained indirectly as linear combinations of the basic comparisons. The across-comparison structural assumptions specify how interaction effects are modeled across treatment comparisons, with independent interactions yielding comparison-specific estimates, common interactions imposing a single pooled interaction across the network, and exchangeable interactions allowing comparison-specific interactions to vary around a shared mean through an additional heterogeneity parameter. Each combination of consistency and across-comparison assumptions corresponds to a distinct parameterization, leading to conceptual challenges in NMR because different combinations of assumptions define models of varying complexity that address related but non-identical hypotheses. Although prior work has described these assumptions conceptually (Dias et al., 2018; Dias, Sutton, et al., 2013; Donegan et al., 2019; Kwarteng et al., 2026), their comparative performance under different scenarios in networks of aggregate data remains underexplored (Proctor et al., 2022).

Beyond considerations of conceptual plausibility, NMR must address practical challenges related to data availability and network structure. Reliable estimation of treatment and interaction effects requires sufficiently connected networks, yet more than a quarter of published networks are sparse, containing limited direct evidence and few studies informing comparisons (Petropoulou et al., 2017). Conventional NMA models encounter difficulties in sparse data settings, which are intensified in NMR, particularly under the independent interaction assumption (Evrenoglou et al., 2024). Although NMR models with a common interaction term across comparisons may numerically mitigate interaction estimation difficulties associated with data sparsity, the consequences of this modeling choice remain unclear. Multi-arm studies are a common data structure in treatment networks (Nikolakopoulou et al., 2014; Petropoulou et al., 2017) and require methods capable of incorporating correlations between observed estimates within multi-arm studies (Rücker et al., 2017; Seide et al., 2020). However, NMR models without standard interaction consistency assumptions may incorporate different parameterizations due to the choice of study-specific reference treatment in the presence of multi-arm studies in NMA and NMR, potentially affecting parameter estimation and results interpretation (Dias, Welton, et al., 2013; Donegan et al., 2019; Spineli, 2022). Because interpretability and practicality of model selection depend on both clinical relevance and practical feasibility (Kwarteng et al., 2026), clear methodological guidance is needed to support appropriate model selection and robust application of NMR.

This article aims to provide methodological guidance on NMR model selection and to clarify the trade-offs associated with alternative NMR modeling assumptions. To this end, we conduct an extensive simulation study in which data are generated under varying conditions, including study design, network density, between-study heterogeneity, and true model specification. The combination of these data-generating settings results in a total of 120 simulation scenarios. By comparing model performance, this work clarifies the consequences of modeling choices and supports applied researchers in selecting appropriate NMR specifications. The remainder of the article is organized as follows. In Section 2, we review the standard network meta-analysis (NMA) model and describe NMR models with independent and common interaction terms, under both imposed and relaxed consistency assumptions. In Section 3, we present the simulation design, and in Section 4 we summarize the main findings. Finally, in Section 5, we discuss the implications of these results and provide practical guidance to support real-world clinical applications of NMR.

## 2. Models

### 2.1. Network Meta-Analysis (NMA)

Consider a network comprising $T$ treatments in $N$ studies, with a common reference treatment across all $N$ studies in the network and any two treatments in the network, denoted $k$ and $q$ , $(k, q = 1,2, \dots, T$ and $k \neq q$). Let $y_{i,hk}$ denote the observed relative effect of treatment $k_{(i)}$ compared with the study-specific reference treatment $h_i$ in study $i = 1,2, \dots, N$ where $k, h_{(i)} \in K_i$, and $K_i$ denotes the total number of arms in study $i$. For simplicity, the subscript $i$ is omitted from the treatment identifiers in subsequent notation. The random effects NMA model can then be parametrized as (Chaimani et al., 2019; Nikolakopoulou et al., 2020),

$$\boldsymbol{y} \sim \mathbf{N}(\boldsymbol{\delta}, \mathbf{V}) \\ \boldsymbol{\delta} \sim \mathbf{N}(\mathbf{X}\boldsymbol{\mu}, \boldsymbol{\Sigma}) \tag{1}$$

where $\mathbf{y} = [y_{i,hk}]' i = 1,2, \dots, N$ represents a vector of all study-specific treatment effects in the network (Kwarteng et al., 2026). The vector $\boldsymbol{\mu}$ represents comparison specific treatment effects, and $\boldsymbol{\delta}$ represents the study-specific treatment effect estimates. Two block-diagonal variance-covariance matrices, $\mathbf{V}$ and $\boldsymbol{\Sigma}$, represent the within and across study variance-covariance matrices, respectively. For a study $i$ in the network, the diagonal elements of each matrix $\mathbf{V_i}$ correspond to the known study-specific variances $\sigma^2_{i,hk}$, with off-diagonals in multi-arm studies corresponding to the respective study-specific covariances between treatments. By convention, we assume a single heterogeneity parameter $\tau^2$ across all pairwise comparisons in the network, which are contained in the diagonal elements of the matrix $\boldsymbol{\Sigma_i}$, while the off-diagonal elements for multi-arm studies are set to $\frac{\tau^2}{2}$ (Higgins & Whitehead, 1996; Lu & Ades, 2009). Due to the assumption of treatment effect consistency, the comparison specific relative effect between any two treatments $k$ and $q$ can be obtained using the treatment effect consistency equation.

$$\mu_{kq} = \mu_{hq} - \mu_{hk} \tag{2}$$

The reference treatment is mathematically arbitrary, and allows the model to estimate $T - 1$ basic parameters, whereas the remaining relative $\binom{T}{2} - (T - 1)$ treatment effects are calculated using the consistency equation (2). The design matrix $\mathbf{X}$, incorporates the consistency equations and has dimension equal to $\sum_{i=1}^{N}(K_i - 1) \times (T - 1)$ with rows corresponding to unique observed pairwise comparisons within each study and columns to the basic NMA parameters, $\mu_{hk}$ (Kwarteng et al., 2026; Salanti et al., 2008). The NMR model fit can rely on either the Bayesian or frequentist framework (Dias et al., 2018; Lu & Ades, 2004; Rücker, 2012). Within a frequentist setting, the basic parameters can be estimated using weighted least squares:

$$\hat{\boldsymbol{\mu}} = \left(\mathbf{X}'\left(\mathbf{V} + \hat{\boldsymbol{\Sigma}}\right)^{-1}\mathbf{X}\right)^{-1} \mathbf{X}'\left(\mathbf{V} + \hat{\boldsymbol{\Sigma}}\right)^{-1}\mathbf{y} \tag{3}$$

The corresponding variance is defined as

$$var(\hat{\boldsymbol{\mu}}) = \left(\mathbf{X}'\left(\mathbf{V} + \hat{\boldsymbol{\Sigma}}\right)^{-1}\mathbf{X}\right)^{-1} \tag{4}$$

and the estimated between studies variance-covariances matrix, $\widehat{\boldsymbol{\Sigma}}$, can be obtained using restricted maximum likelihood (REML), the generalized Der-Simonian Laird estimator, or other techniques (DerSimonian & Laird, 1986; Veroniki et al., 2016; Viechtbauer, 2005). The standard NMA model, defined by both equations (1) and (2) is suitable when no study-level effect modification is present in the network.

### 2.2. Network Meta-Regression

In the presence of effect modifiers, the NMA model in equations (1) and (2) will not explain the between-study heterogeneity (Cooper et al., 2009; Phillippo et al., 2020). NMR extends the NMA model to account for observed study-level effect modifiers and can be parameterized as,

$$\boldsymbol{y} \sim \mathbf{N}(\boldsymbol{\delta}, \mathbf{V}) \qquad (5)$$
$$\boldsymbol{\delta} \sim \mathbf{N}(\mathbf{X}^{*}\boldsymbol{\theta}, \boldsymbol{\Sigma})$$

$\boldsymbol{y}, \boldsymbol{\Sigma}$, and $\mathbf{V}$ are defined as in Section 2.1 and estimation is also possible using the weighted least squares approach. The new vector of unknown parameters $\boldsymbol{\theta}$ consists of the adjusted treatment effect estimates, $\boldsymbol{\mu}$, and the treatment-by-covariate interactions, $\boldsymbol{\beta}$, hereafter termed interactions. The vector of unknown parameters is then,

$$\boldsymbol{\theta} = [\boldsymbol{\mu}, \boldsymbol{\beta}]' \qquad (6)$$

Similar to matrix $\mathbf{X}$, the augmented design matrix $\mathbf{X}^{*}$ for the NMR has in total $\sum_{i=1}^{N}(K_i - 1)$ rows and columns equal to the length of the vector of unknown parameters $\boldsymbol{\theta}$. As in NMA, treatment effect consistency is still imposed using equation (2), thus $\boldsymbol{\mu}$ remains a vector of length $T - 1$. The columns of the augmented design matrix $\mathbf{X}^{*}$ and vectors of unknown parameters $\boldsymbol{\theta}$ and $\boldsymbol{\beta}$ depend on the underlying NMR assumptions. The NMR model fit can rely on either the frequentist or the Bayesian framework (Dias et al., 2018; Kwarteng et al., 2026). Within a frequentist framework, least-squares estimates can be obtained using weighted least squares analogous to equation (3), but incorporating the new design matrix, variance covariance matrix, and vector of unknown parameters

$$\widehat{\boldsymbol{\theta}} = \left(\mathbf{X}^{*\prime}\left(\mathbf{V} + \widehat{\boldsymbol{\Sigma}^{*}}\right)^{-\mathbf{1}}\mathbf{X}^{*}\right)^{-1} \mathbf{X}^{*\prime}\left(\mathbf{V} + \widehat{\boldsymbol{\Sigma}^{*}}\right)^{-\mathbf{1}}\mathbf{y} \qquad (7)$$

with $var(\widehat{\boldsymbol{\theta}}) = \left(\mathbf{X}'\left(\mathbf{V} + \widehat{\boldsymbol{\Sigma}^{*}}\right)^{-\mathbf{1}}\mathbf{X}\right)^{-1}$ and $\widehat{\boldsymbol{\Sigma}^{*}}$, obtainable using techniques available in NMA such as REML, among other approaches (DerSimonian & Laird, 1986; Veroniki et al., 2016; Viechtbauer, 2005).

The corresponding Bayesian NMR implementation requires prior distributions for the unknown parameters $\boldsymbol{\theta}$ and $\tau$. In the absence of strong evidence for priors, non-informative normal priors are commonly specified for $\boldsymbol{\theta}$ (e.g. $\mathrm{N}(0,100^2)$), while weakly informative priors are typically used for the heterogeneity parameter $\tau$ (Röver et al., 2021).

Similar to treatment effect consistency, interaction consistency can enable the estimation of interaction terms through indirect evidence using the interaction consistency equation,

$$\beta_{kq} = \beta_{hq} - \beta_{hk} \tag{8}$$

For models with consistent interaction effects (CIE), the interaction consistency equation is embedded in the augmented design matrix $\mathbf{X}^*$. However, in cases where the assumption of consistency for the interaction terms is weak, investigators might prefer not to impose the additional assumption of consistency for the interaction terms. Alternative NMR models that assume consistent NMA treatment effects but unrelated mean interaction effects (UMIE), estimate interaction parameters only from directly compared treatments. UMIE models do not contain a network-wide reference treatment for the interactions. Instead, reference treatments should be carefully defined on a study-specific basis to ensure that all studies comparing any two treatments $k$ and $q$ are synthesized in an interpretable direction. Directionality can be defined based on study-specific treatment attributes using an indicator variable $Z_{i,kq}$ describing whether treatment $k$ is 'favored' over treatment $q$ in a study (Chaimani & Salanti, 2012; Kwarteng et al., 2026).

The directionality variable is incorporated in the construction of the design matrix $\mathbf{X}^*$ of UMIE models, contributing to the estimation and, later, interpretation of interactions. Multi-arm studies pose similar challenges for UMIE models as they do for unrelated mean effect (UME) models and may result in different parameterizations based on the designated study-specific reference treatment determined from the directionality variable (Dias et al., 2010; Kwarteng et al., 2026).

**2.2.1.Network Meta-Regression with Common and Consistent Interaction Effects (NMR-CCIE)**

Common interaction terms assume the covariate has a similar impact across all treatment comparisons. In NMR models with consistency and common across-comparison interactions (NMR-CCIE), the single basic interaction parameter in equation (6) is estimated for all comparisons relative to a common reference treatment. Due to the interaction consistency equation (8) the non-reference containing interactions are thus zero, as $\beta_{kq} = \beta_{hq} - \beta_{hk} = \beta_{CCIE} - \beta_{CCIE} = 0$. Unlike NMA, the selection of the reference treatment is no longer mathematically arbitrary, and implementations which lack sufficient data for the reference treatment may result in biased estimates or non-estimation. This structure is most appropriate when clinical reasoning supports a uniform covariate effect across all comparisons relative to a reference treatment. The NMR-CCIE model results in a design matrix of dimension $\sum_{i=1}^{N}(K_i - 1) \times (T)$, which may lead to stable and converging interaction estimation in sparse networks, provided that the reference treatment has adequate data. However, in the presence of comparison specific effect modification, the NMR-CCIE model may mask comparison specific variation, resulting in biased estimates.

### 2.2.2. Network Meta-Regression with Independent and Consistent Interaction Effects (NMR-ICIE)

Independent interactions estimate a unique interaction for each pairwise comparison of treatments, such that the interaction term in equation (6) is uniquely estimated for each comparison, $k$ and $q$. NMR models assuming both independent across-comparison structures and consistency equations form NMR-ICIE models. Similar to NMA, models that impose interaction consistency require the definition of a network-wide reference treatment. The NMR-ICIE model estimates $T - 1$ basic interaction parameters relative to the selected reference treatment, and the remaining parameters are estimated using the interaction consistency equation (8). Consequently, NMR-ICIE yields a vector of unknown parameters with length equal to $2(T - 1)$ and a design matrix $\mathbf{X}^*$ of dimension $\sum_{i=1}^{N}(K_i - 1) \times 2(T - 1)$. Reliable estimation in the NMR-ICIE model requires adequate variation in the observed covariate for the compared treatments. When variation is limited, estimates may become less stable. In extreme cases, such as when an observed effect modifier for a treatment is observed in only a single study, variation is then absent and the associated treatment-specific interactions are not identifiable and cannot be estimated.

### 2.2.3. Network Meta-Regression with Independent and Unrelated Mean Interaction Effects (NMR-IUMIE)

UMIE models with independent across-comparison structures (NMR-IUMIE) estimate a unique interaction for each pairwise comparison of treatments, such that the interaction term in equation (6) is uniquely estimated for each comparison. Without the interaction consistency equation, the NMR-IUMIE model, attempts to estimate up to $\binom{T}{2}$ terms for $\boldsymbol{\beta}$. Accordingly, in a fully connected and dense network, the design matrix has dimensions $\sum_{i=1}^{N}(K_i - 1) \times \left((T - 1) + \binom{T}{2}\right)$. However, only comparisons with sufficient direct evidence are estimable, thus this model may suffer in the presence of sparse networks. Furthermore, the required incorporation of a directionality parameter in UMIE models leads to complications of parametrization which may lead to biased or non-estimation of interactions and treatment effects in the presence of multi-arm studies (Dias et al., 2010; Kwarteng et al., 2026).

### 2.2.4. Network Meta-Regression with Common and Unrelated Mean Interaction Effects (NMR-CUMIE)

UMIE models with common across-comparison assumptions (NMR-CUMIE) assume a single interaction in equation (6), which applies to all comparisons and does not impose consistency. The design matrix has the same dimensions as the NMR-CCIE model. However, the two models differ in their consistency assumption. Without consistency, reference treatments are defined for each study and require a directionality parameter as shown. This structure is most appropriate when clinical reasoning supports a uniform covariate effect across all comparisons and a clinically meaningful or known directionality exists. As a UMIE model, the NMR-CUMIE model is also subject to multi-arm parameterization complications and may perform poorly in the presence of multi-arm studies. As a model with common across-comparison structures, the NMR-CUMIE model may mask comparison specific variation in the presence of comparison specific effect modification, resulting in biased estimates.

# 3. Simulation

We design and evaluate our simulations using the recommendations outlined by Morris et al (Morris et al., 2019).

## 3.1. Aims and Data Generation

The simulations in this study are designed to explore the impact of data-generating mechanisms, varying study designs, heterogeneity, and network density on the estimation of adjusted treatment effects and interactions in NMA and NMR. To explore these aspects, we generated 1000 datasets with binary outcomes while varying the characteristics shown in Table 1, leading to 120 scenarios (Evrenoglou et al., 2022; Yao et al., 2025). Each simulated dataset contained equal numbers of participants across treatment arms within each study. The number of participants per arm was generated from a uniform distribution:

$$n_{ik} \sim \text{Unif}(30,60) \tag{9}$$

Study design was specified as networks of exclusively two-arm studies or a mixture of two-arm and multi-arm studies, termed mixed-arm networks. For scenarios involving mixed-arm study designs, we generated fully connected networks in which approximately 10% of studies were multi-arm, reflecting empirical evidence reported by Petropoulou et al., with the remaining studies as two-arm studies, yielding fully-connected networks with an equal number of studies per comparison for $T$ treatments (Petropoulou et al., 2017).

**Table 1: Overview of characteristics:** Overview of characteristics examined in our simulation. Each combination of characteristics formed a unique scenario, and we generated 1000 datasets for each scenario.

| **Characteristic Type** | **Characteristic Value** |
|---|---|
| Density | Fully connected (100%) |
| | Moderately dense (75%) |
| | Moderate sparse (50%) |
| | Sparse (35%) |
| Between-study Heterogeneity | None ($\tau$ =0) |
| | Moderate ($\tau$ =0.28) |
| | High ($\tau$ =0.5) |
| Study Designs | Two-arm |
| | Mixed-arm (two-arm and multi-arm) |
| Interaction Across-Comparison Assumption | Common |
| | Independent |
| | None |
| Interaction Consistency | Imposed |
| | Not imposed |

Irrespective of study design, networks initially began as fully connected treatment networks comprising seven treatments, with four studies per pairwise comparison (Petropoulou et al., 2017). To vary the network density, we progressively removed direct comparisons from the initially fully-connected network, generating networks containing 100%, 75%, 50%, and 35% of possible direct comparisons. The predefined levels of network density resulted in eight distinct network structures (**Figure 1**). Between-study heterogeneity was pre-specified with values $\tau \in \{0,0.29,0.50\}$ in accordance with published empirical priors, and corresponding to no, moderate, and substantial between-study heterogeneity, respectively (Turner et al., 2015).

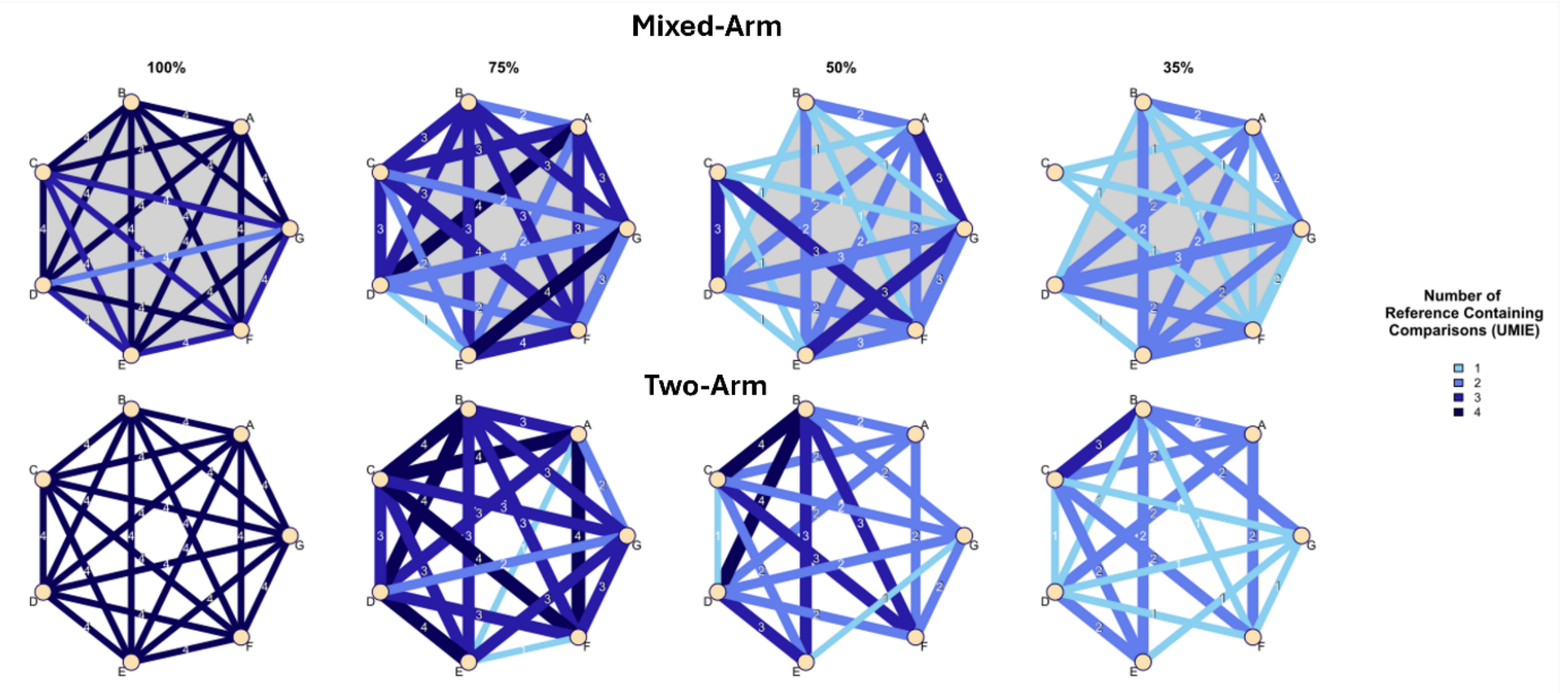

**Figure 1:** Network plots of the scenarios considered in the simulation. Nodes represent treatments and edges represent studies comparing those treatments, with the number of such studies overlaid. Edge colors represent the number of studies considered as direct evidence in the unrelated mean interaction effect models when using treatment order as the directionality parameter. Each network plot corresponds to three between-study heterogeneities and five data generating models

We generated baseline risks of event, $p_{i,h}$, in each study-level reference treatment group $h$, by assuming a binomially distributed $p_{i,h} \sim \text{Binom}(0.30, 0.60)$, for the reference treatment. Note that here we generate this probability regardless of whether $h$ is observed in a specific study. Using the baseline risk, we computed the baseline odds of an event as,

$$odds_{i,h} = \frac{p_{i,h}}{1 - p_{i,h}} \tag{10}$$

The target adjusted treatment effects corresponding to the treatment effects when the effect-modifying covariates are 0, were defined through log-odds ratios ($\log(OR)$s) fixed at equal intervals between 0 and 1 for treatments $h$ through $T$ compared to treatment $h$. That is, for treatment $k$ in a 7-treatment network:

$$\log(OR_{hk}) \, \epsilon \{0,\ 0.17,\ 0.33,\ 0.50,\ \ldots, 1\} \tag{11}$$

Corresponding treatment-pair-specific coefficients $\log(OR)$s then satisfy the NMA consistency equation (2), where $\mu_{kq}$ represents the $\log(OR)$ for two treatments $k$ and $q$.

Between-study heterogeneity was incorporated by generating study-specific treatment effects using a bivariate normal distribution:

$$\begin{bmatrix} \delta_{i,hh} \\ \vdots \\ \delta_{i,hk} \end{bmatrix} \sim N\left( \begin{bmatrix} \log(OR_{hh}) \\ \vdots \\ \log(OR_{hk}) \end{bmatrix}, \begin{bmatrix} \tau^2 & \ldots & \tau^2/2 \\ \vdots & \ddots & \vdots \\ \tau^2/2 & \ldots & \tau^2 \end{bmatrix} \right) \tag{12}$$

Each study included a continuous covariate $X_i \sim Unif(-1, 5)$, which could potentially modify the treatment effect. The covariate effect was incorporated through a treatment-pair-specific coefficient $\beta_{kq}$. The structure of $\beta_{kq}$ was scenario-specific and designed to fulfill the assumption criteria with respect to the data generating model, with non-zero values generated using a uniform distribution $\beta_{kq}$ $\sim$Unif$(-0.75,0.75)$. The $odds_{i,h}$ of event for the non-reference treatments are then calculated as (Evrenoglou et al., 2022),

$$odds_{i,k} = odds_{i,k} * e^{(\delta_{i,hk} + X_i \beta_{hk})}, k \neq h \tag{13}$$

Using the odds, we calculate event probabilities as,

$$p_{i,k} = \frac{odds_{i,k}}{1 + odds_{i,k}} \tag{14}$$

Finally, we generate the number of events for arm $k$ and study $i$, namely $r_{i,k}$ using a Binomial distribution as, $r_{i,k} \sim \text{Binomial}(n_{i,k}, p_{i,k})$.

### 3.1.1. Network Meta-Regression (NMR) Specific Data Generation Considerations

In scenarios assuming consistency, we defined $T - 1$ treatment-pair-specific coefficients, $\beta_{hk}$, for all reference containing comparisons where $k \neq h$. The interaction $\beta_{hh}$ was set to zero. For the CCIE data generating model, $\beta_{hk}$ additionally fulfilled the constraint $\beta_{hk} = \beta_{hq}$. Using the NMR interaction consistency equation, $\beta_{kq}$ was calculated for the remaining non-reference containing comparisons using the consistency equation in equation (8).

For UMIE scenarios, we defined $\frac{T(T-1)}{2}$ treatment-pair-specific coefficients $\beta_{kq} \sim \text{Unif}(-0.75,0.75)$ for each pairwise comparison where $k \neq q$, enabling unrelated interactions across comparisons, although interactions within multi-arm studies remain inherently consistent. Interactions for pairs were defined such that $k$ always preceded $q$ alpha-numerically within each pair. For the CUMIE data generating model, $\beta_{kq}$ additionally fulfilled the constraint $\beta_{hk} = \beta_{kq}$, for all comparisons where $q$ alphanumerically succeeded $k$.

### 3.1.2. Network Meta-Analysis Specific Data Generation Considerations

For data generated in accordance with the NMA model, $\beta_{hk}$ was set to zero for all comparisons, reducing equation (14) to

$$odds_{i,k} = odds_{i,h} * e^{(logOR_{hk})}, k \neq h \tag{15}$$

## 3.2. Evaluated Methods

For each of the 1000 datasets in each scenario, we implemented the five models, NMA, NMR-IUMIE, NMR-CUMIE, NMR-ICIE, and NMR-CCIE, in both frequentist and Bayesian contexts. Directionality parameters in the UMIE models were defined using the alpha-numeric order of the treatments regardless of data generating mechanism. In line with the UMIE data-generating mechanisms, where consistency is expected when the effect-modifying covariate equals zero, model results were evaluated at a covariate value of zero.

For the frequentist models, we used the `netmetareg()` function from the R package **netmeta** and estimated the between-study heterogeneity using restricted maximum likelihood (REML). `netmetareg()` does not require a centered covariate for proper estimation, although centering is often recommended to improve interpretation. For the simulation, covariates were not centered in frequentist implementations because the default scale yields treatment effects corresponding to the zero value of the effect-modifying covariate, as specified in the data-generating model.

We fit the Bayesian models using JAGS for Markov Chain Monte-Carlo (MCMC) sampling via the R package **JAGSUI** (Kellner, 2021). The Bayesian models were fitted based on non-informative priors $\mathrm{N}(0,100^2)$ for the main effects and interaction terms and weakly-informative half-normal priors with scale parameter equal to 1 for the between-study heterogeneity parameter. The models were run with two parallel chains, each containing 10000 burn-in period and 30000 iterations. To ensure proper chain mixing, covariates were centered before implementation in the Bayesian settings (Dias, Sutton, et al., 2013). Bayesian estimates were transformed to correspond to an effect modifier value of zero, thereby ensuring comparability with the frequentist models and the data-generating mechanisms, for which consistency was defined at zero. However, this implementation requirement may introduce bias in the UMIE models. The syntax with detailed information on the Bayesian model implementations is available in section 3 of the supplementary material.

### 3.3. Estimands and Performance Measures

The estimands of interest are the treatment effects, interaction effects, and between-study heterogeneity ($\tau$) on the $\log(OR)$ scale. Performance measures were also evaluated on the $\log(OR)$ scale. Estimates and measures of uncertainty from each model were used to quantify absolute bias, mean squared error (MSE), and coverage.

We calculated the absolute bias for treatment effects, interaction effects, and between-study heterogeneity, by taking the difference between the point estimates and the true value defined in data generation (El Badisy, 2024; Morris et al., 2019). Bias and variance of estimates were used to compute the mean squared error (Morris et al., 2019). Coverage probabilities for treatment and interaction effects were calculated using 95% confidence and credible intervals for the frequentist and Bayesian models, respectively. Coverage probabilities for the interactions reflect the observed proportion of confidence intervals that contained the true interaction for each comparison in each dataset. Performance measures were averaged across treatment comparisons. Coverage probabilities were assessed accounting for Monte Carlo sampling variability, with nominal coverage defined as falling within the 95% confidence interval of $0.95 \pm 1.96\sqrt{\frac{0.95(1-0.95)}{1000}} = (0.94, 0.96)$ for the target coverage of 0.95 in 1000 replications (Burton et al., 2006). Coverage probabilities below this range indicate under-coverage, whereas coverage above this range indicates excessive coverage.

# 4. Results

We present the results of frequentist implementations in two-arm and mixed-arm networks. Detailed findings for Bayesian implementations are provided in Sections S1 and S2 of the supplementary material. The results of the Bayesian and frequentist models are broadly similar, with some exceptions which are discussed in Section S1 of the supplementary materials. More granular results regarding simulation performance measures are available online at `https://kwartengna.shinyapps.io/NMR_Simulation_Results/`.

## 4.1. Bias

### 4.1.1. Bias of NMA Model

Estimates of the mean bias in treatment effects for the NMA model are presented in Figure 2. The associated heterogeneity bias is displayed in **Error! Reference source not found.**. When data were g enerated under the NMA data generating mechanism, treatment effect bias was negligible for the NMA model across all network structures, with the largest magnitude of 0.01 observed in high heterogeneity, sparsely connected networks. The NMA model exhibited its largest bias of treatment effects when data were generated under the CUMIE mechanism. Across all data-generating scenarios with effect modification, the NMA overestimated treatment effects and between-study heterogeneity, reflecting its inability to accommodate effect modification

### 4.1.2. Bias of NMR-CCIE Model

Figure 2 presents the mean bias in treatment effects and interactions for NMR models, while heterogeneity bias is illustrated in **Error! Reference source not found.**. The NMR-CCIE model e xhibited the highest magnitudes of bias in CUMIE data generating scenarios, reaching magnitudes of 0.32 and 0.20 for treatment effects and interactions, respectively. In UMIE data generating scenarios, the NMR-CCIE model persistently overestimated treatment effects and underestimated interactions. Bias was low when the implemented NMR-CCIE model matched the data generating mechanism, with values of bias below 0.01 for treatment effects and interactions. Treatment effect bias was comparably small when the model was applied to networks generated with the NMA mechanism reaching values of 0.02. In ICIE, IUMIE, and CUMIE data generating scenarios, the NMR-CCIE model overestimated the between study heterogeneity, whereas there was no persistent direction of heterogeneity bias when the NMR-CCIE model was applied to data generated according to the NMA and CCIE data.

.

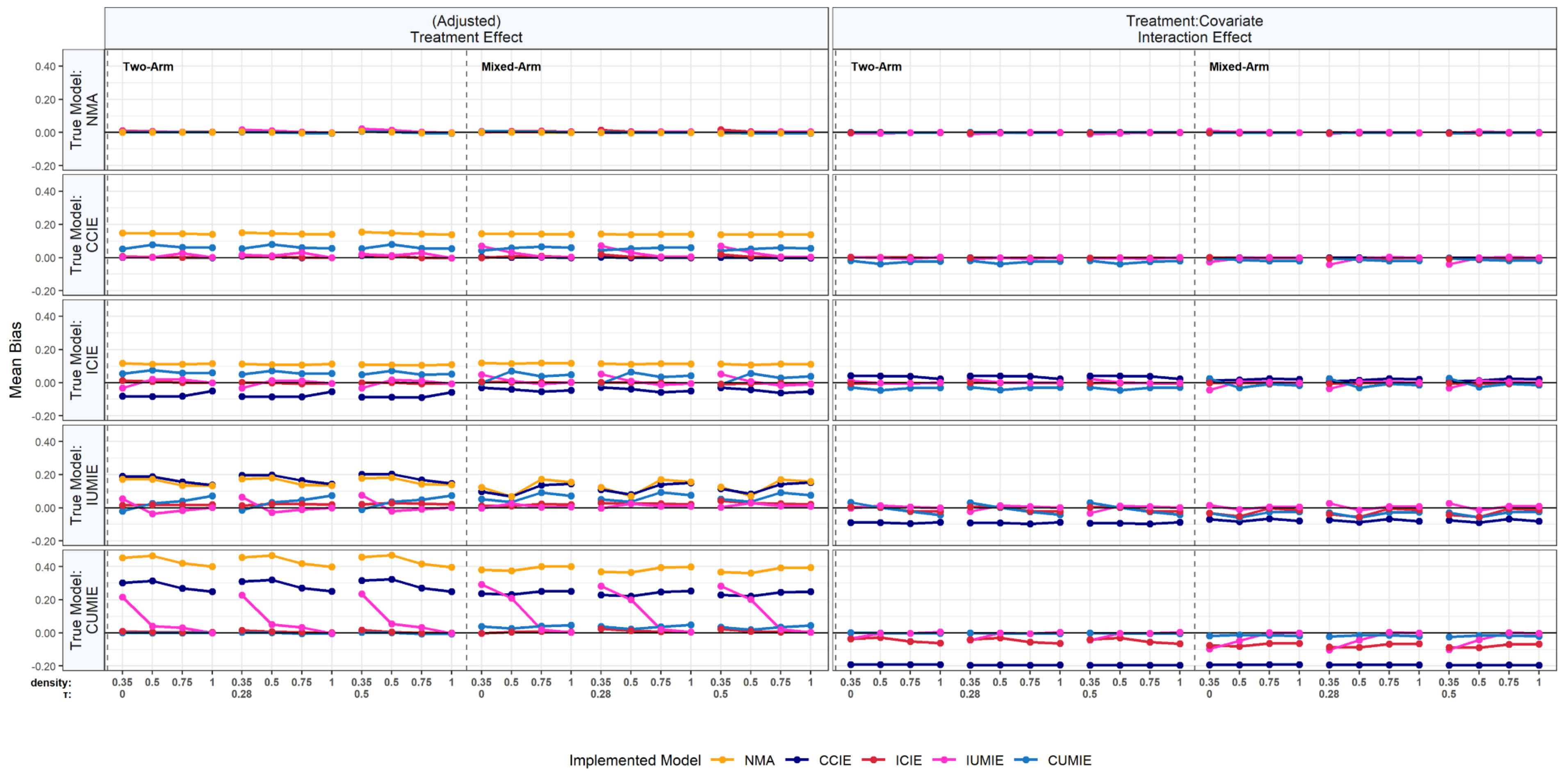

**Figure 2:** Bias Estimates: Bias of relative effect, interaction, and heterogeneity estimates averaged across comparisons for each scenario.

#### 4.1.3. Bias of NMR-ICIE Model

When networks of studies were generated with ICIE interactions, NMR-ICIE generally produced low bias for both treatment effects and interactions, with magnitudes not exceeding 0.01 and 0.01, respectively. Treatment effect bias was comparably small when the model was applied to networks generated with the NMA and CCIE mechanisms with values reaching 0.02 and 0.02 respectively. Bias of interactions was comparably low when the NMR-ICIE model was applied to CCIE generated data reaching values up to 0.01. The NMR-ICIE model showed the greatest bias in treatment effect estimation when data were generated under the IUMIE mechanism, with treatment effect bias values reaching up to 0.04. The highest magnitude of bias for interactions was observed when the NMR-ICIE model was applied to CUMIE generated data with values reaching 0.09. The NMR-ICIE model overestimated heterogeneity in networks generated with UMIE assumptions, whereas there was no persistent direction of heterogeneity bias in networks generated with consistent interactions, or networks generated in accordance with the NMA model.

#### 4.1.4. Bias of NMR-IUMIE Model

For networks generated in accordance with the IUMIE model, the bias of the implemented NMR-IUMIE model reached 0.08 and 0.03 for treatment effects and interactions respectively. In fully connected networks, the NMR-IUMIE model showed relatively low bias for all other data generating models with bias reaching 0.01 and 0.01 for treatment effects and interactions respectively. However, sparser networks showed increased bias, particularly in CUMIE data generating scenarios where bias reached 0.29 for treatment effects and 0.10 for interactions. In networks of mixed-arm studies generated in accordance with the IUMIE and CUMIE models, the NMR-IUMIE model overestimated heterogeneity. However, in networks of two-arm studies, the NMR-IUMIE model showed no persistent direction of heterogeneity estimation.

#### 4.1.5. Bias of NMR-CUMIE Model

The NMR-CUMIE model exhibited the highest bias for treatment effect estimation in networks of mixed-arm studies generated under the IUMIE mechanism, reaching values of up to 0.09. Bias of treatment effects was below 0.01 when the NMR-CUMIE model was applied to NMA-generated data, and was slightly higher in CUMIE-generated networks, reaching values of 0.05. In networks containing multi-arm studies, the NMR-CUMIE model showed persistent overestimation of heterogeneity in all data generating models, except the NMA scenario. However, in networks of two-arm studies, the NMR-CUMIE model only overestimated heterogeneity for data generated under the IUMIE, CCIE, and ICIE mechanisms.

## 4.2. MSE

### 4.2.1. MSE of NMA Model

Patterns of mean squared error across mixed-arm networks are shown in Figure 3. Across all data generating mechanisms, treatment effect MSE from the NMA model increased with increasing between study heterogeneity and decreasing network density. These patterns occurred in both two-arm and mixed-arm networks, indicating strong sensitivity of NMA precision to heterogeneity and sparsity.

### 4.2.2. MSE of NMR-CCIE Model

For the NMR-CCIE model, treatment effect and interaction MSE increased with increasing heterogeneity across all data-generating mechanisms. Treatment effect MSE also increased with decreasing network density in all data generating mechanisms. However, the relationship between interaction MSE and density was less persistent across data generating scenarios. In networks of mixed-arm studies, interaction MSE increased with decreasing density for all data-generating mechanisms. However, in networks of two-arm studies generated under the IUMIE mechanism, this trend in interaction MSE in the NMR-CCIE model did not persist.

### 4.2.3. MSE of NMR-ICIE Model

In mixed-arm networks generated under CCIE, NMA, or IUMIE mechanisms, no clear relationship between MSE and heterogeneity was observed when using the NMR-ICIE model. In contrast, when data were generated under ICIE or CUMIE mechanisms, both treatment effect and interaction MSE increased with increasing heterogeneity in mixed-arm networks. In networks of two-arm studies, treatment effect and interaction MSE increased with heterogeneity across all data-generating mechanisms. In all data generating mechanisms, the treatment effect and interaction MSE from the NMR-ICIE model increased with decreasing density.

### 4.2.4. MSE of NMR-IUMIE Model

Under IUMIE data generation, interaction MSE from the NMR-IUMIE model showed no persistent relationship with heterogeneity. However, when data were generated under ICIE, CCIE, CUMIE, or NMA mechanisms in both two-arm and mixed-arm networks, treatment effect and interaction MSE increased with heterogeneity. For the NMR-IUMIE model, interaction MSE increased with decreasing network density across all data-generating mechanisms. In mixed-arm networks, treatment effect MSE from NMR-IUMIE models also increased as density decreased, whereas no consistent density-related trend was observed in two-arm networks.

.

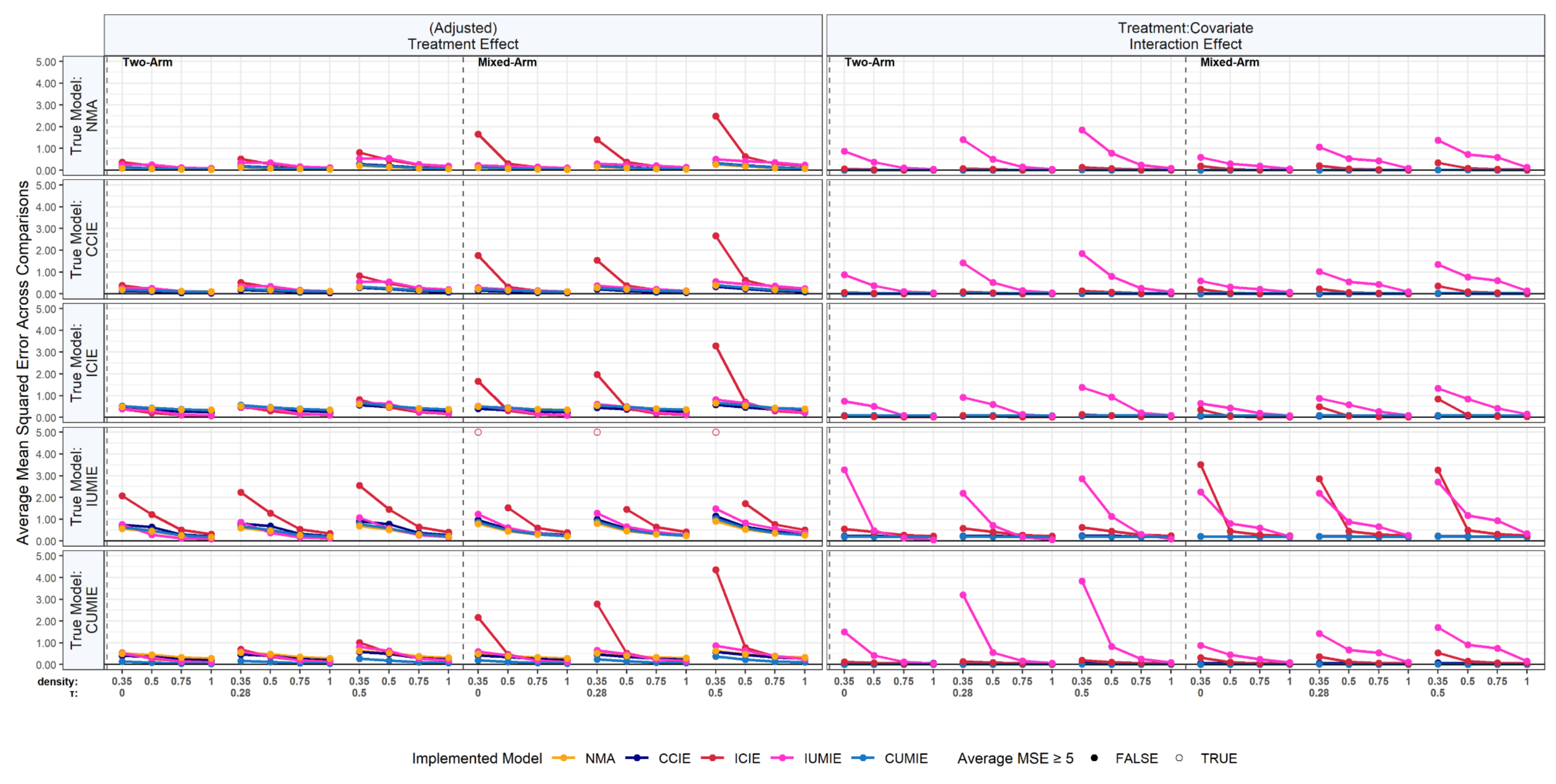

**Figure 3:** MSE: Mean squared error (MSE) of relative effect and interaction estimates averaged across comparisons for each scenario.

#### 4.2.5. MSE of NMR-CUMIE Models

Similar to the NMA model, the treatment effect and interaction MSE from the NMR-CUMIE model increased with increasing between study heterogeneity and decreasing network density in all data generating mechanisms. These patterns were consistent across both two-arm and mixed-arm networks, indicating similarly strong sensitivity of NMR-CUMIE precision to heterogeneity and sparsity

### 4.3. Coverage

#### 4.3.1. Coverage of NMA Model

Coverage patterns varied across models and data-generating mechanisms (Figure 4). The NMA model generally achieved or exceeded nominal coverage for treatment effects when networks were generated without effect modification. Under its own data-generating mechanism, the NMA model only exhibited slight under-coverage of 0.94 in sparse networks of two-arm studies with high heterogeneity. Once effect modification was introduced using CCIE, ICIE, and CUMIE assumptions, the model showed under-coverage of treatment effects across scenarios, and coverage values remained below 0.86. In IUMIE-generated networks lacking full connectivity, the NMA model exhibited nominal or excessive coverage for treatment effects. However, because NMA models are inherently unable to capture effect modification, the coverage of NMA in IUMIE generated data remains inadequate.

#### 4.3.2. Coverage of NMR-CCIE Model

When data were generated under NMA or CCIE mechanisms, the NMR-CCIE models generally maintained nominal treatment-effect coverage, with exceptions in sparse two-arm networks with high heterogeneity (Figure 4). Under the same data generating mechanisms, the NMR-CCIE model showed persistent excess coverage of interactions, indicating conservative confidence intervals. In contrast, under ICIE and CUMIE data generating scenarios, the NMR-CCIE model exhibited persistent under-coverage for both the treatment effects and interactions, and coverage values remained below 0.88. In IUMIE data generating scenarios, the NMR-CCIE model exhibited persistent under-coverage of interactions below 0.21, undermining the range of treatment effect coverage values achieved in the IUMIE data generating scenarios encompassing under-coverage with values of 0.93 and over-coverage with values of 0.97.

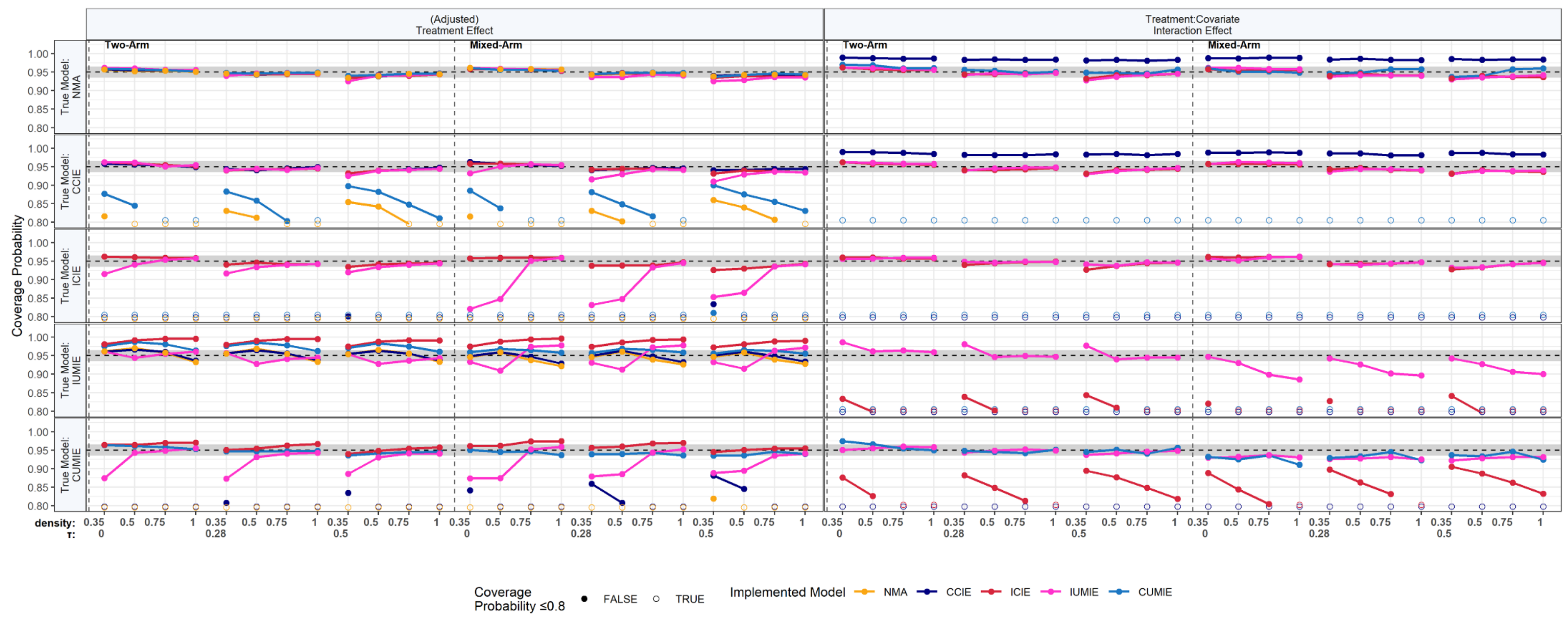

**Figure 4:** Coverage Probabilities: Coverage probability of 95% confidence intervals for the relative effect and interaction parameters averaged across comparisons for each scenario

### 4.3.3. Coverage of NMR-ICIE Model

In networks generated in accordance with the ICIE, CCIE, and NMA mechanisms, the NMR-ICIE models resulted in nominal coverage levels for treatment effects and interactions in networks with low to moderate heterogeneity, but occasionally displayed under-coverage in networks with high between-study heterogeneity (Figure 4). The under-coverage of the NMR-ICIE models in networks with high between study heterogeneity, was more frequently observed in networks containing multi-arm studies, where the median coverage probability was 0.94. In CUMIE and IUMIE data generation scenarios, the NMR-ICIE exhibited nominal or excessive levels of coverage for treatment effects ranging from 0.94 to 1.00, but interactions persistently exhibited under-coverage below 0.91 across scenarios.

### 4.3.4. Coverage of NMR-IUMIE Model

In networks of two-arm studies generated under the NMA or CCIE assumptions, the NMR-IUMIE model achieved nominal coverage levels for both treatment effects and interactions in all scenarios except the high-heterogeneity, low-density scenario (Figure 4). Coverage deteriorated in mixed-arm networks, with treatment effect coverage ranging from 0.91 to 0.96. Treatment effect under-coverage was observed in networks with mixed-arm studies and high heterogeneity, where the median coverage was 0.93 and networks with low density where the median coverage was 0.93. Under the ICIE data generating mechanism, the NMR-IUMIE model maintained nominal coverage in most networks of two-arm studies, but exhibited under-coverage in low-density settings irrespective of heterogeneity and in moderately sparse settings with moderate to high heterogeneity. In mixed-arm ICIE networks, nominal coverage of treatment effects and interactions was observed in fully connected networks, with under-coverage emerging in sparser networks. In networks of two-arm studies, the NMR-IUMIE model exhibited nominal to excess levels of coverage in most IUMIE data generating scenarios, with exceptions in moderately sparse networks with high and moderate between study-heterogeneity. The model also achieved nominal coverage in most CUMIE-generated scenarios of two-arm networks, with exceptions in sparse networks regardless of heterogeneity, as well as in moderately sparse networks with heterogeneity. In IUMIE and CUMIE data generation mechanisms, NMR-IUMIE interaction under-coverage occurred frequently in mixed-arm networks, with coverage probabilities ranging from 0.89 to 0.93, highlighting sensitivity to multi-arm study designs.

### 4.3.5. Coverage of NMR-CUMIE Model

The NMR-CUMIE model retained nominal treatment effect coverage in networks with the NMA data generating mechanism irrespective of study design, with excess interaction coverage in sparse two-arm networks without heterogeneity (Figure 4). In the ICIE and CCIE data generation scenarios, the NMR-CUMIE model persistently showed under-coverage for both treatment effects and interactions, with values below 0.90. In IUMIE data generating scenarios, the NMR-CUMIE model exhibited persistent under-coverage of interactions below 0.43, undermining, the range of treatment effect coverage probabilities achieved in several of the corresponding IUMIE data-generating scenarios from 0.96 to 0.99. In its own data-generating setting, the NMR-CUMIE model achieved nominal or excessive coverage in two-arm networks, except under sparse, high-heterogeneity conditions, where the coverage probability was 0.94. In networks of mixed-arm studies, the NMR-CUMIE model exhibited under-coverage of interactions in the majority of scenarios generated with its corresponding data generating mechanism, indicating susceptibility to multi-arm studies.

# 5. Discussion

## 5.1. Overall Findings and Implications for Practice

In this manuscript we examined the performance of multiple NMR models across several data-generating mechanisms to evaluate the impact of consistency, interaction structure, network density, and between-study heterogeneity on model performance. Our results may provide general guidance for selecting NMR models under the data-generating conditions examined in Figure 5, indicating the practical feasibility of each model in the different settings. Although the NMA model demonstrated adequate coverage in networks without interactions, it exhibited persistent biased treatment effect estimates in the presence of study-level effect modifiers leading to sub-nominal coverage, reflecting the limitations of neglecting effect modification across heterogeneous studies.

| Data Generating Characteristic | Implemented Model: NMA | CCIE | ICIE | IUMIE | CUMIE |
|---|---|---|---|---|---|
| Model: NMA[1] | ✓ | ✓ | ✓ | ✓ | ✓ |
| Model: CCIE[1] | × | ✓ | ✓ | ✓ | × |
| Model: ICIE[1] | × | × | ✓ | ✓ | × |
| Model: IUMIE[1] | × | × | × | ✓ | × |
| Model: CUMIE[1] | × | × | × | ✓ | ✓ |
| Sparse Network[2] | ✓ | ✓ | × | × | ✓ |
| Multi-Arm Network[3] | ✓ | ✓ | ✓ | × | × |
| High Heterogeneity[1] | ⚠ | ⚠ | ⚠ | ⚠ | ⚠ |

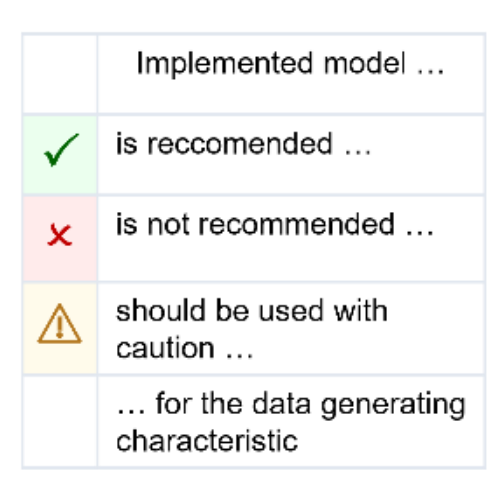

[1] in dense networks of two-arm studies
[2] in conjunction with the reccomended data generating models and networks of two-arm studies
[3] in conjunction with reccomended data generating models and dense networks

**Figure 5:** Overview of characteristic specific model recommendations. Primarily based on the coverage probabilities, this figure indicates when each model could capture the true treatment effects and treatment-covariate interactions in the presence of a particular data generating characteristic.

The NMR-ICIE model generally exhibited small bias in networks with consistent interactions resulting in nominal coverage, regardless of whether the data generating across-comparison interaction were independent or common across comparisons. However, the model remained vulnerable to sparse networks with between-study heterogeneity, highlighting a practical challenge related to data availability. The common across-comparison assumption of the NMR-CCIE model may provide a practical solution for NMR implementation in sparse networks, as the NMR-CCIE models achieved nominal or excessive treatment effect coverage levels when data were generated under their corresponding mechanisms, and were generally robust to between-study heterogeneity and sparsity. Corresponding, confidence intervals for their interactions in NMR-CCIE models were persistently conservative and exhibited excessive coverage. Additionally, though robust to sparse networks, NMR-CCIE models require careful selection of reference treatments for interpretation and warrant future treatment network updates to explore patterns of effect modification at the comparison-specific level.

The coverage of the NMR-IUMIE model captured the true estimates across all effect modifying structures in densely connected networks of two-arm studies, yet coverage declined in mixed-arm or sparse networks, highlighting practical limits of the model's generalizability. The NMR-CUMIE model exhibited persistent sub-nominal coverage in mixed-arm networks, again illustrating the limits of UMIE models in the presence of multi-arm studies. In the presence of multi-arm studies, models with the consistency assumption may be preferable, because they can incorporate all comparisons of a multi-arm study.

In the absence of multi-arm studies and between-study heterogeneity, all NMR models, including the NMR-CCIE and NMR-CUMIE models, are capable of approximating the NMA model in scenarios without effect modification. This property occurs because all NMR models effectively reduce to the NMA model when interaction parameters approach zero and are correctly estimated. Furthermore, though the CCIE, ICIE, and IUMIE models can be parameterized to estimate similar comparisons, in the presence of effect modification, the CUMIE model is distinct from the CCIE and ICIE models and does not approximate the same relationships due to the different underlying comparisons which are then aggregated in their modeling structures.

### 5.2. Strengths and Limitations

This work provides a detailed exploration of the impact of network density on NMR performance in the presence of known data-generating mechanisms. To our knowledge, this is the most extensive simulation study evaluating several network meta-regression models in a wide range of data-generating settings. Our findings are consistent with existing simulation studies that examined the ability of models to capture treatment effect estimation (Harari et al., 2023; Kiefer et al., 2020; Proctor et al., 2022; Seide et al., 2020). However, we extend these simulation studies by exploring the interaction estimation capabilities of several NMR modelling assumptions (Proctor et al., 2022).

In applied settings, the true underlying data generation mechanism is unknown. Furthermore, tools for assessing the plausibility of consistency assumptions are limited. In densely-connected networks of two-arm studies, the NMR-IUMIE model can provide insight by estimating independent interactions for each comparison, but this approach is insufficient in multi-arm and sparse networks. Thus, UMIE models could benefit from further adaptations to appropriately incorporate multi-arm studies, as has been performed for the UME equivalent (Spineli, 2022). Future research is also needed for developing model-selection frameworks and diagnostic tools that help analysts identify appropriate consistency and effect-modification assumptions, particularly in clinical settings where the true generating mechanism is uncertain.

### 5.3. Conclusion

Our findings highlight the importance of aligning model assumptions with both the anticipated form of effect modification and the structure of the evidence network. In the presence of effect modification, the use of NMA models, which do not account for treatment-covariate interactions, can lead to biased estimates of both treatment effects and between-study heterogeneity. In such settings, NMR models should be considered, and the choice of model should be guided by a combination of statistical rationale and clinical expertise.

In practice, NMR models that assume consistent interactions provide practical benefits, particularly in networks that include multi-arm studies. In densely connected networks, the NMR-IUMIE model can serve as a tool for assessing the assumptions of other NMR models in well-connected networks of two-arm studies, but its applicability in multi-arm and sparse structures remains challenging, thus it is insufficient for exploring assumptions in many contexts. Sparse networks present additional challenges for NMR irrespective of consistency assumptions and often motivate the use of models assuming a common interaction term. When adopting models with consistent interactions, the plausibility of the consistency assumption should be assessed, while models assuming a common interaction require scrutiny of the across-comparison assumption (Donegan et al., 2019).

# References

Ades, A. E., Welton, N. J., Dias, S., Phillippo, D. M., & Caldwell, D. M. (2024). Twenty years of network meta-analysis: Continuing controversies and recent developments. *Research Synthesis Methods*, jrsm.1700. https://doi.org/10.1002/jrsm.1700

Borenstein, M. (2009). *Introduction to meta-analysis*. Wiley.

Burton, A., Altman, D. G., Royston, P., & Holder, R. L. (2006). The design of simulation studies in medical statistics. *Statistics in Medicine*, *25*(24), 4279–4292. https://doi.org/10.1002/sim.2673

Caldwell, D. M., Welton, N. J., & Ades, A. E. (2010). Mixed treatment comparison analysis provides internally coherent treatment effect estimates based on overviews of reviews and can reveal inconsistency. *Journal of Clinical Epidemiology*, *63*(8), 875–882. https://doi.org/10.1016/j.jclinepi.2009.08.025

Chaimani, A. (2015). Accounting for baseline differences in meta-analysis. *Evidence Based Mental Health*, *18*(1), 23–26. https://doi.org/10.1136/eb-2014-102035

Chaimani, A., Caldwell, D. M., Li, T., Higgins, J. P., & Salanti, G. (2019). 11. Undertaking network meta-analyses. In J. P. T. Higgins, J. Thomas, J. Chandler, M. Cumpston, T. Li, M. J. Page, & V. A. Welch (Eds.), *Cochrane Handbook for Systematic Reviews of Interventions* (1st ed., pp. 285–320). Wiley. https://doi.org/10.1002/9781119536604.ch11

Chaimani, A., & Salanti, G. (2012). Using network meta-analysis to evaluate the existence of small-study effects in a network of interventions. *Research Synthesis Methods*, *3*(2), 161–176. https://doi.org/10.1002/jrsm.57

Cooper, N. J., Sutton, A. J., Morris, D., Ades, A. E., & Welton, N. J. (2009). Addressing between-study heterogeneity and inconsistency in mixed treatment comparisons: Application to stroke prevention treatments in individuals with non-rheumatic atrial fibrillation. *Statistics in Medicine*, *28*(14), 1861–1881. https://doi.org/10.1002/sim.3594

Deeks, J. J., Higgins, J. P., Altman, D. G., & on behalf of the Cochrane Statistical Methods Group. (2019). Analysing data and undertaking meta-analyses. In J. P. T. Higgins, J. Thomas, J. Chandler, M. Cumpston, T. Li, M. J. Page, & V. A. Welch (Eds.), *Cochrane Handbook for*

*Systematic Reviews of Interventions* (1st ed., pp. 241–284). Wiley. https://doi.org/10.1002/9781119536604.ch10

DerSimonian, R., & Laird, N. (1986). Meta-analysis in clinical trials. *Controlled Clinical Trials*, *7*(3), 177–188. https://doi.org/10.1016/0197-2456(86)90046-2

Dias, S., Ades, A. E., Welton, N. J., Jansen, J. P., & Sutton, A. J. (2018). *Network meta-analysis for Decision Making* (1st ed.). John Wiley & Sons, Inc. https://learning.oreilly.com/library/view/-/9781118647509/?ar

Dias, S., Sutton, A. J., Welton, N. J., & Ades, A. E. (2013). Evidence synthesis for decision making 3: Heterogeneity—Subgroups, meta-regression, bias, and bias-adjustment. *Medical Decision Making : An International Journal of the Society for Medical Decision Making*, *33*(5), 618–640. https://doi.org/10.1177/0272989X13485157

Dias, S., Welton, N. J., Marinho, V. C. C., Salanti, G., Higgins, J. P. T., & Ades, A. E. (2010). Estimation and Adjustment of Bias in Randomized Evidence by Using Mixed Treatment Comparison Meta-Analysis. *Journal of the Royal Statistical Society Series A: Statistics in Society*, *173*(3), 613–629. https://doi.org/10.1111/j.1467-985X.2010.00639.x

Dias, S., Welton, N. J., Sutton, A. J., Caldwell, D. M., Lu, G., & Ades, A. E. (2013). Evidence Synthesis for Decision Making 4: Inconsistency in Networks of Evidence Based on Randomized Controlled Trials. *Medical Decision Making*, *33*(5), 641–656. https://doi.org/10.1177/0272989X12455847

Donegan, S., Dias, S., & Welton, N. J. (2019). Assessing the consistency assumptions underlying network meta-regression using aggregate data. *Research Synthesis Methods*, *10*(2), 207–224. https://doi.org/10.1002/jrsm.1327

Donegan, S., Welton, N. J., Tudur Smith, C., D'Alessandro, U., & Dias, S. (2017). Network meta-analysis including treatment by covariate interactions: Consistency can vary across covariate values. *Research Synthesis Methods*, *8*(4), 485–495. https://doi.org/10.1002/jrsm.1257

Donegan, S., Williams, L., Dias, S., Tudur-Smith, C., & Welton, N. (2015). Exploring Treatment by Covariate Interactions Using Subgroup Analysis and Meta-Regression in Cochrane Reviews: A

Review of Recent Practice. *PLOS ONE*, *10*(6), e0128804. https://doi.org/10.1371/journal.pone.0128804

El Badisy, I. (2024). *mcstatsim: Monte Carlo Statistical Simulation Tools Using a Functional Approach* [Computer software]. https://CRAN.R-project.org/package=mcstatsim

Evrenoglou, T., Metelli, S., Thomas, J., Siafis, S., Turner, R. M., Leucht, S., & Chaimani, A. (2024). Sharing information across patient subgroups to draw conclusions from sparse treatment networks. *Biometrical Journal*, *66*(3), 2200316. https://doi.org/10.1002/bimj.202200316

Evrenoglou, T., White, I. R., Afach, S., Mavridis, D., & Chaimani, A. (2022). Network meta-analysis of rare events using penalized likelihood regression. *Statistics in Medicine*, *41*(26), 5203–5219. https://doi.org/10.1002/sim.9562

Harari, O., Soltanifar, M., Cappelleri, J. C., Verhoek, A., Ouwens, M., Daly, C., & Heeg, B. (2023). Network meta-interpolation: Effect modification adjustment in network meta-analysis using subgroup analyses. *Research Synthesis Methods*, *14*(2), 211–233. https://doi.org/10.1002/jrsm.1608

Higgins, J. P. T., Thomas, J., Chandler, J., Cumpston, M., Li, T., Page, M. J., & Welch, V. A. (Eds.). (2019). 10.11 Investigating heterogeneity. In *Cochrane Handbook for Systematic Reviews of Interventions* (1st ed.). Wiley. https://doi.org/10.1002/9781119536604

Higgins, J. P. T., & Whitehead, A. (1996). Borrowing strength from external trials in a meta-analysis. *Statistics in Medicine*, *15*(24), 2733–2749. https://doi.org/10.1002/(SICI)1097-0258(19961230)15:24%3C2733::AID-SIM562%3E3.0.CO;2-0

Kellner, K. (2021). *jagsUI: A Wrapper Around "rjags" to Streamline "JAGS" Analyses*. https://CRAN.R-project.org/package=jagsUI

Kiefer, C., Sturtz, S., & Bender, R. (2020). A simulation study to compare different estimation approaches for network meta-analysis and corresponding methods to evaluate the consistency assumption. *BMC Medical Research Methodology*, *20*(1), 36. https://doi.org/10.1186/s12874-020-0917-3

Kwarteng, N., Evrenoglou, T., Mueller, J., Elsaesser, M., Schramm, E., Schwarzer, G., & Nikolakopoulou, A. (2026). *Illustrating the Assumptions of Meta-Regression in Treatment Networks*. In Review. https://doi.org/10.21203/rs.3.rs-8235913/v1

Lu, G., & Ades, A. (2009). Modeling between-trial variance structure in mixed treatment comparisons. *Biostatistics (Oxford, England)*, *10*(4), 792–805. https://doi.org/10.1093/biostatistics/kxp032

Lu, G., & Ades, A. E. (2004). Combination of direct and indirect evidence in mixed treatment comparisons. *Statistics in Medicine*, *23*(20), 3105–3124. https://doi.org/10.1002/sim.1875

Lumley, T. (2002). Network meta-analysis for indirect treatment comparisons. *Statistics in Medicine*, *21*(16), 2313–2324. https://doi.org/10.1002/sim.1201

Morris, T. P., White, I. R., & Crowther, M. J. (2019). Using simulation studies to evaluate statistical methods. *Statistics in Medicine*, *38*(11), 2074–2102. https://doi.org/doi.org/10.1002/sim.8086

Nikolakopoulou, A., Chaimani, A., Veroniki, A. A., Vasiliadis, H. S., Schmid, C. H., & Salanti, G. (2014). Characteristics of Networks of Interventions: A Description of a Database of 186 Published Networks. *PLoS ONE*, *9*(1), e86754. https://doi.org/10.1371/journal.pone.0086754

Nikolakopoulou, A., White, I. R., & Salanti, G. (2020). Network Meta-Analysis. In C. H. Schmid, T. Stijnen, & I. R. White, *Handbook of Meta-Analysis* (1st ed., pp. 187–218). Chapman and Hall/CRC. https://doi.org/10.1201/9781315119403-10

Nixon, R. M., Bansback, N., & Brennan, A. (2007). Using mixed treatment comparisons and meta-regression to perform indirect comparisons to estimate the efficacy of biologic treatments in rheumatoid arthritis. *Statistics in Medicine*, *26*(6), 1237–1254. https://doi.org/10.1002/sim.2624

Petropoulou, M., Nikolakopoulou, A., Veroniki, A.-A., Rios, P., Vafaei, A., Zarin, W., Giannatsi, M., Sullivan, S., Tricco, A. C., Chaimani, A., Egger, M., & Salanti, G. (2017). Bibliographic study showed improving statistical methodology of network meta-analyses published between 1999 and 2015. *Journal of Clinical Epidemiology*, *82*, 20–28. https://doi.org/10.1016/j.jclinepi.2016.11.002

Phillippo, D. M., Dias, S., Ades, A. E., Belger, M., Brnabic, A., Schacht, A., Saure, D., Kadziola, Z., & Welton, N. J. (2020). Multilevel network meta-regression for population-adjusted treatment comparisons. *Journal of the Royal Statistical Society. Series A, (Statistics in Society)*, *183*(3), 1189–1210. https://doi.org/10.1111/rssa.12579

Polanin, J. R., & Williams, R. T. (2016). Overcoming obstacles in obtaining individual participant data for meta-analysis. *Research Synthesis Methods*, *7*(3), 333–341. https://doi.org/10.1002/jrsm.1208

Proctor, T., Zimmermann, S., Seide, S., & Kieser, M. (2022). A comparison of methods for enriching network meta-analyses in the absence of individual patient data. *Research Synthesis Methods*, *13*(6), 745–759. https://doi.org/10.1002/jrsm.1568

Riley, R. D., Lambert, P. C., Staessen, J. A., Wang, J., Gueyffier, F., Thijs, L., & Boutitie, F. (2008). Meta-analysis of continuous outcomes combining individual patient data and aggregate data. *Statistics in Medicine*, *27*(11), 1870–1893. https://doi.org/10.1002/sim.3165

Röver, C., Bender, R., Dias, S., Schmid, C. H., Schmidli, H., Sturtz, S., Weber, S., & Friede, T. (2021). On weakly informative prior distributions for the heterogeneity parameter in Bayesian random-effects meta-analysis. *Research Synthesis Methods*, *12*(4), 448–474. https://doi.org/10.1002/jrsm.1475

Rücker, G. (2012). Network meta-analysis, electrical networks and graph theory. *Research Synthesis Methods*, *3*(4), 312–324. https://doi.org/10.1002/jrsm.1058

Rücker, G., Cates, C. J., & Schwarzer, G. (2017). Methods for including information from multi-arm trials in pairwise meta-analysis. *Research Synthesis Methods*, *8*(4), 392–403. https://doi.org/10.1002/jrsm.1259

Salanti, G. (2012). Indirect and mixed-treatment comparison, network, or multiple-treatments meta-analysis: Many names, many benefits, many concerns for the next generation evidence synthesis tool. *Research Synthesis Methods*, *3*(2), 80–97. https://doi.org/10.1002/jrsm.1037

Salanti, G., Higgins, J. P. T., Ades, A. E., & Ioannidis, J. P. A. (2008). Evaluation of networks of randomized trials. *Statistical Methods in Medical Research*, *17*(3), 279–301. https://doi.org/10.1177/0962280207080643

Schmid, C. H., Stark, P. C., Berlin, J. A., Landais, P., & Lau, J. (2004). Meta-regression detected associations between heterogeneous treatment effects and study-level, but not patient-level, factors. *Journal of Clinical Epidemiology*, *57*(7), 683–697. https://doi.org/10.1016/j.jclinepi.2003.12.001

Seide, S. E., Jensen, K., & Kieser, M. (2020). A comparison of Bayesian and frequentist methods in random-effects network meta-analysis of binary data. *Research Synthesis Methods*, *11*(3), 363–378. https://doi.org/10.1002/jrsm.1397

Spineli, L. M. (2022). A Revised Framework to Evaluate the Consistency Assumption Globally in a Network of Interventions. *Medical Decision Making*, *42*(5), 637–648. https://doi.org/10.1177/0272989X211068005

Thompson, S. G., & Sharp, S. J. (1999). Explaining heterogeneity in meta-analysis: A comparison of methods. *Statistics in Medicine*, *18*(20), 2693–2708. https://doi.org/10.1002/(SICI)1097-0258(19991030)18:20%3C2693::AID-SIM235%3E3.0.CO;2-V

Turner, R. M., Jackson, D., Wei, Y., Thompson, S. G., & Higgins, J. P. T. (2015). Predictive distributions for between-study heterogeneity and simple methods for their application in Bayesian meta-analysis. *Statistics in Medicine*, *34*(6), 984–998. https://doi.org/10.1002/sim.6381

Van Der Worp, H., Holtman, G. A., & Blanker, M. H. (2023). Challenges In Performing An Individual Participant–level Data Meta-analysis. *European Urology Focus*, *9*(5), 705–707. https://doi.org/10.1016/j.euf.2023.10.012

Veroniki, A. A., Jackson, D., Viechtbauer, W., Bender, R., Bowden, J., Knapp, G., Kuss, O., Higgins, J. P. T., Langan, D., & Salanti, G. (2016). Methods to estimate the between-study variance and its uncertainty in meta-analysis. *Research Synthesis Methods*, *7*(1), 55–79. https://doi.org/10.1002/jrsm.1164

Viechtbauer, W. (2005). Bias and Efficiency of Meta-Analytic Variance Estimators in the Random-Effects Model. *Journal of Educational and Behavioral Statistics*, *30*(3), 261–293. https://doi.org/10.3102/10769986030003261

White, I. R., Barrett, J. K., Jackson, D., & Higgins, J. P. T. (2012). Consistency and inconsistency in network meta-analysis: Model estimation using multivariate meta-regression. *Research Synthesis Methods*, *3*(2), 111–125. https://doi.org/10.1002/jrsm.1045

Yao, M., Mei, F., Zou, K., Li, L., & Sun, X. (2025). Comparison of Prior Distributions for the Heterogeneity Parameter in a Rare Events Meta-Analysis of a Few Studies. *Pharmaceutical Statistics*, *24*(2), e2448. https://doi.org/10.1002/pst.2448

## Declarations

**Funding:** This project was supported by the Deutsche Forschungsgemeinschaft (DFG, German Research Foundation) – Project-ID 499552394 – SFB 1597. TE was supported by the Deutsche Forschungsgemeinschaft (DFG, German Research Foundation) under Project ID 554095932. AN has received funding from the European Research Executive Agency under the HORIZON TMA MSCA Postdoctoral Fellowships - European Fellowships grant agreement 101244226.